\documentclass[onecolumn]{pasj00}
\SetRunningHead{Y. Kawano et al.}{Galaxy Structures and External
 Perturbations in Gravitational Lenses}
\title{Galaxy Structures and External Perturbations in Gravitational Lenses}
\author{Yozo \textsc{Kawano}\altaffilmark{1},
    Masamune \textsc{Oguri}\altaffilmark{2},
    Takahiko \textsc{Matsubara}\altaffilmark{1},
    and Satoru \textsc{Ikeuchi}\altaffilmark{1}}
\altaffiltext{1}{Department of Physics and Astrophysics, Nagoya University, 
Chikusa-ku, Nagoya 464-8062}
\email{kawano@a.phys.nagoya-u.ac.jp}
\altaffiltext{2}{Department of Physics, School of Science, University of 
Tokyo, Bunkyo-ku, Tokyo 113-0033}
\Received{2003 December 17}
\Accepted{2004 February 12}
\Published{$\langle$publication date$\rangle$}
\begin{document}
\maketitle
\begin{abstract}
In modeling strong gravitational lens systems, one often adopts simple
models, such as singular isothermal elliptical density plus lowest-order
external perturbation. However, such simple models may mislead us if the real
mass distribution is more complicated than that in the assumed models. In
particular, assumptions on mass models are crucial in studying flux ratio
anomalies that have been suggested as evidence for a cold dark-matter
substructure. We reinvestigated four quadruple lens systems using
power-law Fourier models, which have advantages of clear physical
meanings and the applicability of a linear method, as well as a simple
singular isothermal elliptical density model. We also investigated the
effect of external perturbations, including a singular isothermal sphere, 
lowest-order expansion, and next-order expansion. We have found that the
$\cos3\theta$ terms of the primary galaxy and/or of external perturbation
significantly reduce the $\chi^2$ in PG 1115+080 and B1422+231. In
particular, we could reproduce the flux ratios of B1422+231 with
next-order external perturbation assuming 5\% flux uncertainties,
suggesting that external perturbation cannot be described by a simple
singular isothermal sphere approximation. On the other hand, we could not
fit B0712+472 and B2045+265 very well even with our models, although the
$\chi^2$ were reduced compared with the case of using simple models. Our results
 clearly demonstrate that both the primary lens galaxy and the external
 perturbation are often more complicated than we usually assume.   
\end{abstract}

\section{Introduction}
Gravitational lenses of distant quasars are quite useful in cosmological
and astrophysical studies, such as measuring the Hubble constant
from time delays \citep{refsdal64}, studies of galaxy formation and
evolution, and direct probes of galaxy structure, including cold
dark-matter substructures.
The method used to measure the Hubble constant does not rely on
the distance ladder, which inevitably
suffers various systematic errors. After the discovery of the first
lens system, Q 0957+561 \citep{walsh79}, many lens systems have been
studied in detail, adopting various lens models. However, the model
uncertainties, including several intrinsic degeneracies
\citep{falco85,gorenstein88,saha00}, prevent us from accurately deriving
the Hubble constant (e.g.,
\cite{wambsganss94,keetonks97,keetonk97,courbin97,wucknitz99,witt00,chiba02,wucknitz02,oguri02,ogukawa03}).   

Among several applications of gravitational lenses, flux-ratio anomalies
have attracted much attention. There are mainly three explanations
for them: Effects of the interstellar medium, problems in smooth
models, and lensing by the substructure (e.g., \cite{kocda03}). While
substructure lensing is often regarded as being the best explanation for
the flux anomalies (e.g., \cite{mao98,chiba02b}), the former two explanations have
not yet been sufficiently investigated to reach the conclusion that
the flux-ratio anomalies are connected with the substructure. The limitation comes from
the fact that it is difficult to apply all of the models, which are physical
and complicated, to the observed lens systems. In addition, since
galaxies are believed to consist of not only luminous matter, but
also dark matter, the shapes of lens galaxies do not have to be
elliptical, nor other simple shapes, as the isophotes indicate. Therefore,
it is of great importance to investigate the contribution of many forms
of the angular structure of the lens potential to find the true origin of
the flux anomalies. Indeed, \citet{evansw03} 
showed that some of lens systems that have been claimed to be lenses with
flux anomalies  can be well fitted by a smooth lens model using a
power-law Fourier potential (see also \cite{ogukawa03}). Kochanek and
Dalal (2003) studied
the contribution of only the isothermal $m=3$ and $m=4$ multipoles to
the flux ratio anomalies reported so far. We then reinvestigate
some of the systems with not only multipoles, but also several types of
external perturbations.

Our aim in this paper is to discuss the contribution of both multipoles of
the primary lens galaxy and external perturbation of the group, in
particular the $\cos3\theta$ terms that are not usually included in lens
modeling. We consider whether some multipoles can produce an acceptable fit
without substructure or not. We also discuss the quadruple lens
system B0712+472 in detail because of the interesting observations that
the ellipticity of the galaxy and the shear of the group seem to be
significantly misaligned ($\sim \timeform{90D}$) to each other, and that
previous authors estimated the shear to be small compared to
observations using the one-shear model \citep{jackson98,fass02}. Since
more than three images are necessary to constrain the mass models strongly,
we restrict our considerations to four quadruple lens systems (PG 1115+080,
B1422+231, B0712+472, and B2045+265). There are three reasons for this
restriction. First, the existence of a substructure of the lens systems has been claimed
to account for the observed flux ratios (e.g., \cite{kocda03}). Second, there are
no satellite galaxies that complicate the lens model. Third, there
are sufficient astrometric, photometric, and spectroscopic (redshift) data.
In this paper, we consider only isothermal models; however our analysis
can be easily extended to non-isothermal radial mass profiles. 

This paper is organized as follows. In section \ref{sec:modelmethod}, we
describe the models that we used to investigate four systems. In section
\ref{sec:results}, we show how we applied the  models to four systems
and give the results. We finally discuss the results and give
conclusions in section \ref{sec:conc}. The linear method that we used
is presented in Appendix \ref{app:linear}.

\section{Lens Model and Method \label{sec:modelmethod}}

\subsection{Fundamental Equations of Gravitational Lensing}
A source at $\mbox{\boldmath $x$}_{\mathrm{s}}$ and an image $i$ at
$\mbox{\boldmath $x$}_i$ are related through the lens equation,
\begin{eqnarray}
 \mbox{\boldmath $x$}_i - \mbox{\boldmath $x$}_{\mathrm{s}}=
{\bf \nabla}\psi(\mbox{\boldmath $x$}_i),
\label{eq:lenseq}
\end{eqnarray}
where $\psi(\mbox{\boldmath $x$})$ is the projected lens potential.
Thus, the magnification factor is given by
\begin{eqnarray}
 \mu (\mbox{\boldmath $x$}_i)=\left[ {\rm det}\left(\frac{\partial 
\mbox{\boldmath $x$}_{\mathrm{s}}}{\partial \mbox{\boldmath $x$}_i}\right) 
\right]^{-1}.
\end{eqnarray}
We can also calculate the equation of the relative time delay between
image $i$ and $j$ using 
\begin{eqnarray}
 \Delta t_{ij} &=& \frac{1+z_{\mathrm{d}}}{c} \frac{D_{\mathrm{d}} 
D_{\mathrm{s}}}{D_{\mathrm{ds}}} \nonumber\\
&& \times \left[ \frac{1}{2} \Delta \mbox{\boldmath $x$}_i^2 - \frac{1}{2}
\Delta \mbox{\boldmath $x$}_j^2
- \psi(\mbox{\boldmath $x$}_i) + \psi(\mbox{\boldmath $x$}_j)\right],
\label{eq:tij}
\end{eqnarray}
where $z_{\mathrm{d}}$ is the redshift of the lens object and $c$ is the speed of 
light; $D_{\mathrm{d}}$, $D_{\mathrm{s}}$, $D_{\mathrm{ds}}$ are the angular diameter
distances to the lens, to the source, and from the lens to the source,
respectively; $\Delta \mbox{\boldmath $x$}_i=
\mbox{\boldmath $x$}_i - \mbox{\boldmath $x$}_{\mathrm{s}}$
\citep{schneider92}. The two-dimensional Poisson equation 
relates the dimensionless density distribution
$\kappa(\mbox{\boldmath $x$})$ to the lens potential as
\begin{eqnarray}
 \nabla^2 \psi(\mbox{\boldmath $x$})=2\kappa(\mbox{\boldmath $x$}).
\end{eqnarray}

\subsection{Lens Model}
For the potential of the primary lens galaxy, we apply the singular 
isothermal elliptical density (SIED) model and/or the power-law Fourier
potential (PLF; \cite{ogukawa03}; Evans, Witt 2003) model. The
former model may be called the ``standard'' model in the sense that it is
widely used in the mass modeling of a lens galaxy. The two-dimensional
density profile of SIED is given, in the Cartesian coordinates $(x, y)$
centered on the primary lens galaxy, as 
\begin{eqnarray}
 \kappa_{\mathrm{galaxy}}(\mbox{\boldmath $x$})=b(x^2+y^2/q^2)^{-1/2},
\end{eqnarray}
where $b$ is the mass parameter that characterizes the Einstein radius
and $q$ is the axis ratio. The ellipticity, $e$, is given by $1-q$. We
used the latter model, PLF, to investigate the contribution of multipoles,
particularly an $m=3$ multipole, which is neglected in usual lens
modelings of lens systems. In addition, this model has an advantage
that we can apply a linear method (Appendix \ref{app:linear}).
The lens potential of the power-law Fourier model is given by
\begin{eqnarray}
\psi_{\mathrm{galaxy}}=r^{\beta} F(\theta),
\end{eqnarray}
where a Fourier expansion of the angular function $F(\theta)$ is
\begin{eqnarray}
 F(\theta)&=&a_0+\sum_{m=1}^{\infty}(a_m \cos m\theta+b_m \sin m\theta) \\
          &=&a_0+\sum_{m=1}^{\infty} A_m \cos \{m(\theta-\theta_m)\},
\end{eqnarray}
in polar coordinates $(r, \theta)$. The variables $A_m=(a_m^2+b_m^2)^{1/2}$ and
$\theta_m$ are the amplitude and the position angle of the multipole,
respectively. In the present work, we considered only the case of $\beta=1$,
which has a similar radial profile as in SIED,
while the parameter $\beta$ can take values between 0
(point mass) and 2 (mass sheet).

While the influence of the galaxy group which embeds the primary lens
galaxy should be taken into account (e.g., \cite{keetonks97,witt00}),
detailed information on the density distribution of the group is
uncertain in most cases. The potential near the lens galaxies is often
described by a singular isothermal sphere (SIS) or the potential
expanded around the center of the lens galaxy
\citep{kocha91,bernstein99,keeton00} as
\begin{eqnarray}
 \psi_{\mathrm{ex}}&=& \psi_0+\mbox{\boldmath $\psi$}_{1}\cdot \mbox{\boldmath $x$}
\nonumber\\
&&+r^2\left[\frac{1}{2}\kappa_0+\frac{1}{2}\gamma \cos
 2(\theta-\theta_{\gamma})\right] \nonumber\\
&&+r^3\left[\frac{1}{4}\sigma \sin(\theta-\theta_{\sigma})
-\frac{1}{6}\delta \sin 3(\theta-\theta_{\delta})\right]\nonumber\\
&&+{\cal O}(r^4).
\label{eq:expert}
\end{eqnarray}
The first three terms cannot be fully constrained by observables, because of
the lens degeneracies \citep{falco85,gorenstein88,saha00}. Thus, one
usually uses an external shear (XS),
\begin{eqnarray}
 \psi_{\mathrm{xs}}=\frac{\gamma}{2}r^2\cos 2(\theta-\theta_{\gamma})
=\frac{\gamma_1}{2}r^2\cos 2\theta+\frac{\gamma_2}{2}r^2\sin 2\theta,
\end{eqnarray}
where $\gamma_1=\gamma \cos 2\theta_{\gamma}$ and $\gamma_2=\gamma
\sin 2\theta_{\gamma}$.
In the present work, we also added the terms of third order ($r^3$) to
the lens potential in a similar way of e.g., Bernstein and Fischer (1999)
enforcing $\delta=-(3/2)\sigma$ and $\theta_{\delta}=\theta_{\sigma}$
(as for a singular isothermal group). In this case, the explicit
expression of the additional potential (X3) is 
\begin{eqnarray}
 \psi_{\mathrm{x3}}=\frac{r^3}{4}\sigma
\left[\sin (\theta-\theta_{\sigma})+\sin3
  (\theta-\theta_{\sigma})\right].
\label{eq:X3}
\end{eqnarray}

In summary, we used SIS, SIED, and PLF models for the primary
lens galaxy, and XS, SIS, or (XS+X3) models for the group.

\subsection{Method \label{subsec:method}}
We used image positions and flux ratios as constraints on the lens models.
The relative time delays between images were
used not only for determining the Hubble constant, but also
for additional constraints on the lens models. The number of constraints was more
than ten for a quadruple lens system, while the typical number of model
parameters is equal to or more than seven. The goodness of fit was
determined by
$\chi^2$ statistics based on the image positions, the flux ratios, and the
relative time delays: 
\begin{eqnarray}
\chi_{\mathrm{pos}}^2&=&\sum_i \frac{|\mbox{\boldmath $x$}_i-
\mbox{\boldmath $x$}_i^{\mathrm{(model)}}|^2}{\sigma_{xi}^2}, 
\\
\chi_{\mathrm{flux}}^2&=&\sum_{i<j}
    \frac{(f_{ij}-f_{ij}^{\mathrm{(model)}})^2}{\sigma_{fij}^2}, 
\\
\chi_{\mathrm{time}}^2&=&\sum_{i<j} \frac{(\Delta t_{ij}-\Delta
    t_{ij}^{\mathrm{(model)}})^2}{\sigma_{tij}^2}. 
\end{eqnarray}
In the above equations, $\mbox{\boldmath $x$}_i$ are the observed image
positions, $f_{ij}$ are the observed flux ratios, and $\Delta t_{ij}$
are the observed relative time delays. The suffix ``(model)'' indicates
the model predictions for each observable. The uncertainties in the
observables are represented by $\sigma_{xi}$, $\sigma_{fij}$, and
$\sigma_{tij}$. We minimized the total chi-square, $\chi_{\mathrm{
total}}^2=\chi_{\mathrm{pos}}^2+\chi_{\mathrm{flux}}^2+\chi_{\mathrm{time}}^2$
\citep{press92}, to obtain the best-fit parameters,
$\mbox{\boldmath $u$}_{\mathrm{best}}$. In this fitting, (i) we solved
non-linear lens equations to derive the model predictions for the image
positions, (ii) calculated the relative time delays and the flux ratios
from the predicted image positions and the model parameters, and
(iii) evaluated and minimized $\chi_{\mathrm{total}}^2$. Finally, we
evaluated the reduced chi-square, $\chi_{\mathrm{total}}^2/N_{\mathrm{dof}}$, where
$N_{\mathrm{dof}}$ is the number of degrees of freedom. In the case of
the PLF model, the $\chi^2$ minimization was significantly simplified because
of the linearity of the equations (see Appendix \ref{app:linear}).

\section{Results \label{sec:results}}
About 80 gravitational-lens systems have been reported so far. One
third of them are quadruple-lens systems, and the rest are double-lens systems.
However, we considered only quadruple lens systems, because more than six
observables are typically needed to strongly constrain lens models.
Specifically, we analyzed the systems PG 1115+080 \citep{weymann80},
B1422+231 \citep{patnaik92}, B0712+472 \citep{jackson98}, and B2045+265
\citep{fass99}. We adopted a flat lambda-dominated universe with
$(\Omega_0, \Omega_{\Lambda})=(0.3, 0.7)$, where $\Omega_0$ is the
density parameter of matter and $\Omega_{\Lambda}$  is the dimensionless
cosmological constant.  The Hubble constant in units of $100$ km
s$^{-1}$ Mpc$^{-1}$ is denoted by $h$, as usual. 

\subsection{Data}
The positions and fluxes of most of the gravitationally lensed images
are relatively well constrained within 10 mas and about five percent
uncertainties, respectively, while measuring time delays from light
curves are very difficult. For most of the systems, we used the data of
the image positions and the fluxes by the CfA/Arizona Space Telescope Lens
Survey\footnote{http://cfa-www.harvard.edu/castles/}
(CASTLES; \cite{falco99}), and for some of the systems the data was
taken from the papers cited below. The uncertainties of the flux ratios were
slightly broadened (see below) to account for possible systematics, such
as possible microlensing (e.g., \cite{mao98}), extinction, and quasar
variability during a time delay.

The system PG 1115+080 is a QSO ($z_{\mathrm{s}}=1.72$) gravitationally lensed
by an early-type galaxy of $z_{\mathrm{d}}=0.31$, which looks very spherical
\citep{keeton98,iwamuro00}. The lens galaxy belongs to a group of
galaxies. The QSO images consist of four components (fold-type) with the
maximum angular separation of 2.32 arcsec. In order to constrain models, we used
the CASTLES data of the image positions and the fluxes, and
\citet{barkana97}'s analysis of the time delays. The flux uncertainties
were broadened to 20\%. 

The second system, B1422+231, that we analyzed is known as a system with possible
substructure microlensing
\citep{mao98,keeton01,chiba02b,bradac02,keetongp03,kocda03}.
The elongated galaxy ($z_{\mathrm{d}}=0.34$)
associated with a poor group gravitationally lenses a QSO ($z_{\mathrm{
s}}=3.62$), and leads to a cusp-type four-imaged system. The ellipticity,
$e$, of the galaxy is $0.27\pm 0.13$ \citep{impey96}. There are radio data
of this system (\cite{patnaik92}, 1999), having uncertainties of
50 $\mu$arcsec for QSO images and a few percents for their fluxes. Some
other observational properties, such as the effects of the Lyman
$\alpha$ \citep{bechtold95,petry98} and the variability
\citep{yee96,koopmans03}, have been observed. We used the data of
\citet{patnaik99}, assuming flux uncertainties of 5\%. 

We also investigated an interesting system B0712+472 \citep{jackson98}. 
The redshifts of the elongated primary lens galaxy and the source QSO
are $0.41$ and $1.34$, respectively. Interestingly, the primary
lens galaxy is not associated with a group of galaxies, but a
foreground group with a mean redshifts of $0.29$ are overlapped \citep{fass02}.
The light distribution of the galaxy is very elongated with
an ellipticity of between $0.4$ and $0.5$. The positions of the QSO and
the galaxy images that we used were from \citet{jackson98} and CASTLES, respectively.
For the fluxes, radio data of \citet{jackson98} was used,
assuming 20\% uncertainties.

The system B2045+265 was discovered by a radio survey of the Cosmic Lens
All-Sky Survey (CLASS). Five radio components were clearly observed
\citep{fass99}. The four components are the lensed QSO ($z_{\mathrm{s}}=1.28$) and
another component is the radio core of the primary lens galaxy ($z_{\mathrm{
d}}=0.87$), which is associated with a compact group of galaxies. The
optical spectrum of the galaxy is similar to that of nearby Sa galaxies. We
used the data of CASTLES for the galaxy and that of \citet{fass99} for the QSO.
The flux uncertainties were assumed to be 20\% uncertainties.

\subsection{Implications from Models \label{subsec:impl}}
\begin{table}
\begin{center}
\caption{Results of PG 1115+080.}\label{table:1115}
    \begin{tabular}{llll}
\hline\hline
      Model & $\chi^2/N_{\mathrm{dof}}$ & Best-fit parameters & Note\\
\hline
      SIED & 537.9/8 & $e=0.32$, $\theta_{\mathrm{e}}=\timeform{66D}$ & \\
      SIS+XS & 257.0/8 & $\gamma=0.11$, $\theta_{\gamma}=\timeform{-115D}$ & \\
      SIED+XS & (27.4+1.5+1.8)/6 & $e=0.18$, $\theta_{\mathrm{e}}=\timeform{89D}$ & \\
              & & $\gamma=0.09, \theta_{\gamma}=\timeform{-132D}$ & \\
      SIS+(XS+X3) & (0.6+2.9+0.3)/6 & $\gamma=0.12, \theta_{\gamma}=\timeform{-115D}$ & Acceptable fit \\
      & &$\sigma=0.008, \theta_{\sigma}=\timeform{-160D}$ & \\
      SIS+SIS & (2.6+2.7+0.2)/7 & $r_{\mathrm{group}}=\timeform{12''.4},
      \theta_{\mathrm{group}}=\timeform{-116D}$ & Acceptable fit\\
      (m=2)+XS & (27.8+1.5+1.7)/6 &$\gamma=0.11, \theta_{\gamma}=\timeform{-129D}$
       &\\
      && $A_2/a_0=0.03,\theta_2=\timeform{6D}$& \\
      (m=3)+XS & (1.0+3.3+0.7)/6 & $\gamma=0.12, \theta_{\gamma}=\timeform{-115D}$&
      Acceptable fit\\
      && $A_3/a_0=0.002,\theta_3=\timeform{-14D}$& \\
\hline
  \multicolumn{4}{@{}l@{}}{\hbox to 0pt{\parbox{85mm}{\footnotesize
      Notes. The $\chi^2$ is the total chi-square, $\chi_{\mathrm{pos}}^2
      +\chi_{\mathrm{flux}}^2+\chi_{\mathrm{time}}^2$. Model denotes the
      used potential of galaxy+group. The position of the
      group center is $(r_{\mathrm{group}}, \theta_{\mathrm{group}})$.
     }}}
    \end{tabular}
\end{center}
\end{table}

Table \ref{table:1115} gives the results of PG 1115+080 with typical
models that we described in previous section. Some models
[($m=3$)+XS, SIS+(XS+X3), and SIS+SIS] well reproduce the image
positions, the flux ratios, and the relative time delays. While most
previous authors applied a SIED+XS or a SIED+SIS, our results clearly imply that
the lens requires the term $\cos3\theta$, which might be a component
of the primary lens galaxy ($m=3$ multipole), or be that of the external
perturbation of the group (X3). A very small amplitude
($A_3/a_0\approx 0.002$) of the $m=3$ multipole of the galaxy is
consistent with the galaxy light distribution, which is very spherical.
While some of the models including only XS as the external
perturbation fail to reproduce the observables, inclusion of the
contribution from the group represented by SIS and XS+X3 always gives a
better fit, i.e., an X3 perturbation is important to improve the fit.
The best fit of the angles in the external perturbation are
$\theta_{\gamma}=\timeform{-115D}$ and $\theta_{\sigma}=\timeform{-160D}$.
These angles are consistent with optical \citep{impey98} and X-ray
\citep{grant03} observations of the group. Therefore, we conclude that
there is a degeneracy between an $m=3$ multipole of the lens galaxy
and an X3 perturbation of the group in this lens system, possibly
because the galaxy is very spherical and the external shear is
small. The mass distribution of the galaxy might be spherical and
the ellipticity of the galaxy might be {\it unnecessary}.
\citet{schechter97} also used the model of a SIS+SIS, but
they did not allow the galaxy position to vary. Allowing a variable
galaxy position is essential.  

\begin{table}
\begin{center}
\caption{Results of B1422+231.}\label{table:1422}

    \begin{tabular}{llll}
\hline\hline
      Model & $\chi^2/N_{\mathrm{dof}}$ & Best-fit parameters & Note\\
\hline
      SIED+XS & (2.5+12.2)/4 & $e=0.28$, $\theta_{\mathrm{e}}=\timeform{-56D}$& \\
      && $\gamma=0.16$, $\theta_{\gamma}=\timeform{127D}$& \\
      SIED+SIS & (1.5+9.0)/3 &$e=0.22$, $\theta_{\mathrm{e}}=\timeform{-58D}$ &\\
      && $r_{\mathrm{group}}=\timeform{9''.5}$, $\theta_{\mathrm{group}}=\timeform{128D}$&\\
      SIED+(XS+X3) & (0.1+3.8)/2 & $e=0.25$, $\theta_{\mathrm{e}}=\timeform{-59D}$
      &Acceptable fit\\
      && $\gamma=0.20$, $\theta_{\gamma}=\timeform{134D}$& \\
      && $\sigma=0.02$, $\theta_{\sigma}=\timeform{63D}$& \\
      $[$SIED+(m=3)$]$+XS & (0.1+6.1)/2 & $e=0.29$, $\theta_{\mathrm{e}}=\timeform{-55D}$
      &\\
      && $\gamma=0.17$, $\theta_{\gamma}=\timeform{129D}$& \\
      && $A_3/a_0=0.007$, $\theta_3=\timeform{-16D}$\\
      (m=3)+XS & 781.8/4 & $\gamma=0.21$, $\theta_{\gamma}=\timeform{124D}$
      &\\
      && $A_3/a_0=0.01$, $\theta_3=\timeform{-18D}$ & Too large $A_3$\\
\hline
  \multicolumn{4}{@{}l@{}}{\hbox to 0pt{\parbox{85mm}{\footnotesize
      Notes. The $\chi^2$ is the total chi-square, $\chi_{\mathrm{pos}}^2
      +\chi_{\mathrm{flux}}^2$. Model denotes the
      used potential of galaxy+group.
     }}}
    \end{tabular}
\end{center}
\end{table}

The results of B1422+231 are given in table \ref{table:1422}. While the
SIED+XS and even the SIED+SIS cannot reproduce the observables of the
system, particularly the flux ratios, including the perturbation of X3
significantly reduced the $\chi^2/N_{\mathrm{dof}}$ to less than 2 $[$see the
(SIED+$m=3$)+XS model$]$. The fact that the angles of the external
perturbations are somewhat strange,
$|\theta_{\gamma}-\theta_{\sigma}|=|\timeform{134D}-\timeform{63D}|=\timeform{71D}$,
implies that the structure of the group is not so simple as SIS.
Recent X-ray observations \citep{grant03} support this
interpretation. Moreover, the direction of
the $\sigma$ perturbation in our modeling $[$equation (\ref{eq:expert})$]$ coincides
with the direction of the gradient of the group density distribution.
Note that a part of the large misalignment of the external perturbations
($\gamma$ and $\sigma$) may arise from limiting the
amplitudes and the directions of the $r^3$ perturbation
$[$equation (\ref{eq:X3})$]$. 
Since the system was well-fitted with 5\% flux
uncertainties, it may be unnecessary to consider microlensing by
substructure, or other compact structures, for this system, which supports
the analysis of Keeton, Gaudi, and Petters (2003) using the cusp relation. 

\begin{table}
\begin{center}
\caption{Results of B0712+472.}\label{table:0712}

    \begin{tabular}{llll}
\hline\hline
      Model & $\chi^2/N_{\mathrm{dof}}$ & Best-fit parameters & Note\\
\hline
      SIED & (33.5+54.6)/6 & $e=0.33$, $\theta_{\mathrm{e}}=\timeform{50D}$&\\
      SIS+XS & (53.2+86.8)/6 & $\gamma=0.12$, $\theta_{\gamma}=\timeform{49D}$&\\
      SIED+XS & (10.1+3.0)/4 & $e=0.73$, $\theta_{\mathrm{e}}=\timeform{56D}$& \\
      &&$\gamma=0.33$, $\theta_{\gamma}=\timeform{-34D}$\\
      SIED+SIS & (9.3+4.9)/3  &$e=0.61$, $\theta_{\mathrm{e}}=\timeform{55D}$& \\
      && $r_{\mathrm{group}}=\timeform{6''.0}$, $\theta_{\mathrm{group}}=\timeform{152D}$\\
      SIED+(XS+X3) & (2.5+4.8)/2 & $e=0.69$, $\theta_{\mathrm{e}}=\timeform{56D}$
      & \\
      && $\gamma=0.32$, $\theta_{\gamma}=\timeform{-43D}$& \\
      && $\sigma=0.19$, $\theta_{\sigma}=\timeform{123D}$& \\
      $[$SIED+(m=3)$]$+XS & (3.9+0.7)/2 & $e=0.63$, $\theta_{\mathrm{e}}=\timeform{54D}$
      & Too large $A_3$\\
      && $\gamma=0.22$, $\theta_{\gamma}=\timeform{-39D}$& \\
      && $A_3/a_0=0.01$, $\theta_3=\timeform{37D}$ & \\
\hline
  \multicolumn{4}{@{}l@{}}{\hbox to 0pt{\parbox{85mm}{\footnotesize
      Notes. The $\chi^2$ is the total chi-square, $\chi_{\mathrm{pos}}^2
      +\chi_{\mathrm{flux}}^2$. Model denotes the
      used potential of galaxy+group.
     }}}
    \end{tabular}
\end{center}
\end{table}

We also show results of B0712+472 in table \ref{table:0712}. In this
system, it does not seem that fitting by the SIED+XS model was significantly
improved by including other additional terms, such as an $m=3$ multipole
of the galaxy and X3 terms of the group. Therefore, it may be better to
consider the existence of substructure(s) in this galaxy. However, we
note that the curves of the flux ratios are variable, and perhaps further  
observational studies are needed (see \cite{koopmans03}). Another
interesting result is that both the ellipticity of the primary lens
galaxy and the amplitude of the external shear of the foreground
group are large in a {\it two}-shear model, such as SIED+XS.
This is in marked contrast with the {\it one}-shear model studied so far  
\citep{keeton98,jackson98,fass02} in which the ellipticity of the light
is larger than that of the mass and the external shear is very small
($\gamma<0.1$). We believe that the results with ellipticity plus
external shear are more realistic, since the two shears are in good
agreement with that of observed light. The conflict between the two
can be easily understood by the fact
$|\theta_{\mathrm{e}}-\theta_{\gamma}|\approx \timeform{90D}$, where $\theta_{\mathrm{e}}$ is
the angle of the mass ellipticity. Here, we roughly represent the
potential of the SIED as the PLF to $m=2$,
\begin{eqnarray}
 \psi_{\rm SIED}\approx r\left[ a_0+A_{\mathrm{e}} \cos 2(\theta-\theta_{\mathrm{e}}\right)],
\end{eqnarray}
where $A_{\mathrm{e}}$ is the amplitude of the mass ellipticity.
Their contribution to the lens potential
is approximately expressed (without $r$ dependences) as
$A_{\mathrm{e}}\cos2(\theta-\theta_{\mathrm{e}})+(\gamma/2)\cos2(\theta-\theta_{\gamma})
=A_{\mathrm{e}}\cos2(\theta-\theta_{\mathrm{e}})+(\gamma/2)\cos2(\theta-\theta_{\mathrm{e}}-\timeform{90D})=(A_{\mathrm{e}}-\gamma/2)\cos2(\theta-\theta_{\mathrm{e}})$.
Therefore, while two amplitudes, $A_{\mathrm{e}}$ and $\gamma$, of ellipticity-only
and external shear-only models are small, those of realistic ellipticity
plus external shear model are large.

\begin{table}
\begin{center}
\caption{Results of B2045+265.}\label{table:2045}
    \begin{tabular}{llll}
\hline\hline
      Model & $\chi^2/N_{\mathrm{dof}}$ & Best-fit parameters& Note\\
\hline
      SIED+XS & 481.1/4 & $e=0.19$, $\theta_{\mathrm{e}}=\timeform{-70D}$&\\
      && $\gamma=0.10$, $\theta_{\gamma}=\timeform{-69D}$\\
      SIED+SIS & (16.0+48.0)/3 & $r_{\mathrm{sis}}=\timeform{1''.5},
      \theta_{\mathrm{sis}}=\timeform{-66D}$&Not smooth\\
      SIED+(XS+X3) & (2.3+15.2)/2 & $e=0.81$, $\theta_{\mathrm{e}}=\timeform{-68D}$
      & \\
      && $\gamma=0.66$, $\theta_{\gamma}=\timeform{21D}$\\
      && $\sigma=0.21$, $\theta_{\gamma}=\timeform{160D}$\\
      $[$SIED+(m=3)$]$+XS & 311.4/2 & $e=0.37$, $\theta_{\mathrm{e}}=\timeform{24D}$
      &\\
      && $\gamma=0.22$, $\theta_{\gamma}=\timeform{-67D}$\\
      && $A_3/a_0=0.03$, $\theta_3=\timeform{-70D}$ \\
\hline
  \multicolumn{4}{@{}l@{}}{\hbox to 0pt{\parbox{85mm}{\footnotesize
      Notes. The $\chi^2$ is the total chi-square, $\chi_{\mathrm{pos}}^2
      +\chi_{\mathrm{flux}}^2$. Model denotes the
      used potential of galaxy+group.
     }}}
    \end{tabular}
\end{center}
\end{table}

In the case of B2045+265, all models that we considered could not reproduce 
the observables well (table \ref{table:2045}), particularly the flux ratios.
Some models reduce the $\chi^2$, but we find that they are not physical.
However, the system shows very large (up to $\sim 40 \%$) extrinsic
variations of fluxes \citep{koopmans03} on time scales of several months,
depending on the frequency \citep{fass99}. Therefore, it may be
premature to conclude that this system shows substructure lensing, given
that the nature of their variabilities still remains to be understood. One
possibility is the Galactic scintillation as \citet{koopmans03}
indicated, and another one is anomalous gravitational lensing by the
inner structure of the primary lens galaxy. 

\section{Conclusion \label{sec:conc}}
We investigated four quadruple gravitational lens system (PG
1115+080, B1422+231, B0712+472, and B2045+265) using mass models of SIED
or PLF as well as external perturbations described by SIS, or several 
truncations of the expansion of the external perturbation. We
performed $\chi^2$ fittings with all of the models presented in section
\ref{sec:modelmethod} in order to study the contribution of terms that
are usually not included in mass modelings, such as $m=3$ multipoles of
primary lens galaxy and next-order external perturbations.

We have found that the $m=3$ multipoles and/or X3 external perturbation
significantly reduce the $\chi^2$ in PG 1115+080 and B1422+231. In PG
1115+080, not only the external perturbation of the $r^3$ order (X3), but
also an internal $m=3$ multipole, reduces the $\chi^2$ to statistically
acceptable values, but the ellipticity of the galaxy does not. Their
amplitudes and directions seem to be physical, comparing the results with 
optical and X-ray observations of the galaxy and the group
\citep{impey98,grant03}; in a ($m=3$)+XS model the ratio $A_3/a_0\approx
0.002$ is small enough to reproduce the observed very spherical galaxy,
and in a SIS+(XS+X3) model the ratio $\sigma/\gamma \approx 0.1$ and the
angles $\theta_{\gamma}=\timeform{-115D}$ and $\theta_{\sigma}=\timeform{-160D}$
can be expected from the positions of the
member galaxies and the X-ray brightness contours. A SIS group
also provides an acceptable fit. A significant (but small) $m=3$
galaxy multipole might be induced by interactions with the member galaxies.
Regarding B1422+231, several authors concluded that the ``anomalous'' flux
ratios reflected the existence of substructure lensing (e.g.,
\cite{mao98,chiba02b,bradac02,kocda03}). However, Keeton, Gaudi, and
Petters (2003) found that the cusp relation does not reveal a significant anomaly in
the system. In fact, our analysis has revealed that a SIED+(XS+X3)  
model without substructure reproduces the fluxes within 5\% errors. 
In addition, the X-ray brightness contours (figure 2 of  \cite{grant03})
imply that the direction of the gradient of the group mass is $\approx
\timeform{90D}-\timeform{180D}$, and that the shapes of the contours are significantly
deformed, which are consistent with the results of SIED+(XS+X3) model.
In the other systems, B0712+472
and B2045+265, the contribution of the $m=3$ multipoles and X3 external
perturbations to the lens potentials do not
significantly improve the model fitting. In B0712+472, however, we find
that the ellipticity and shear are well misaligned, $\theta_{\mathrm{e}}\approx
\timeform{56D}$ and $\theta_{\gamma}\approx \timeform{146D}$, and that the
result is very consistent with the lens galaxy and the foreground
group observed by Fassnacht and Lubin (2002). Therefore, the models including only
ellipticity or shear may underestimate these (see subsection \ref{subsec:impl}).
The relatively large shear may arise from the elongated shape of the
group.

In summary, the terms of an $m=3$ multipole of the primary lens galaxy
and of an X3 external perturbation, which are often neglected in strong
lens analyses, sometimes improve the fit significantly. The
contribution of the group particularly needs to be studied in more
detail. Simple models (in particular one-shear models) may often
mislead us.

\bigskip
This work is supported by a Grant-in-Aid for JSPS Fellows.

\appendix

\section{The Linear Method \label{app:linear}}
\begin{figure}
 \begin{center}
   \FigureFile(70mm,50mm){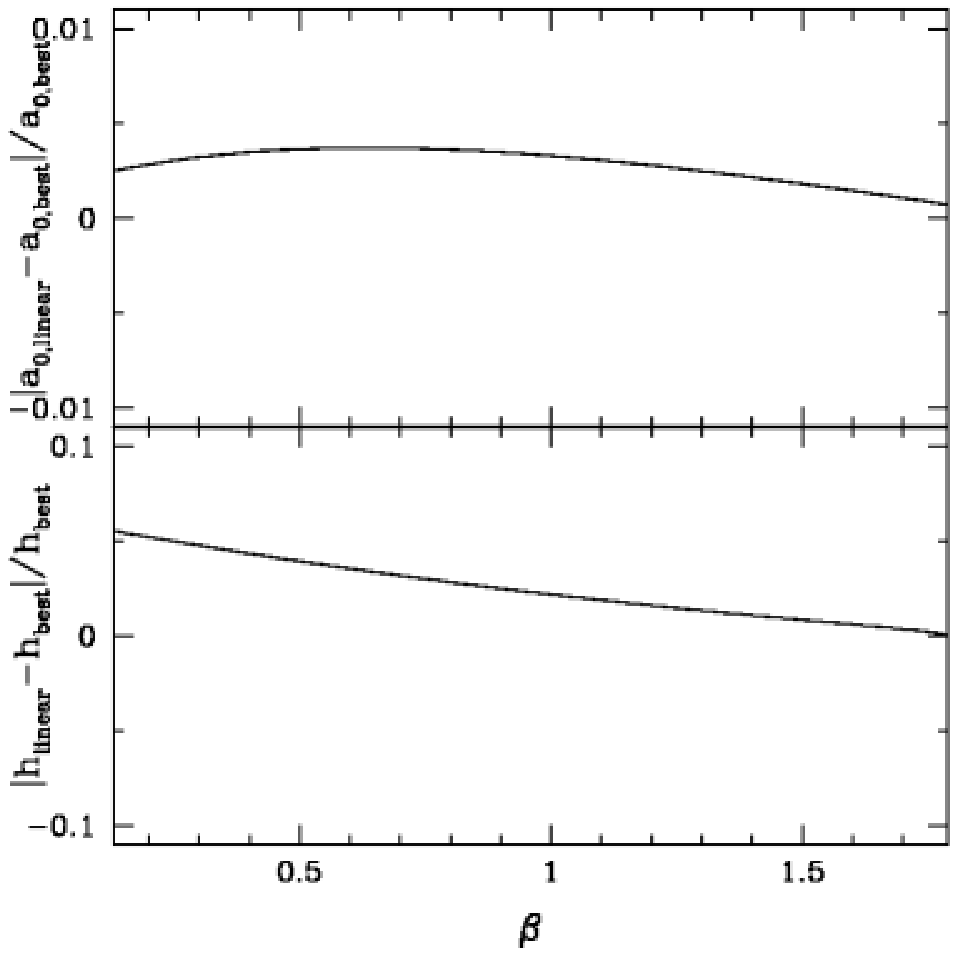}
 \end{center}
 \caption{The normalized deviations of $a_0$ and $h$ of 
$\mbox{\boldmath $u$}_{\mathrm{linear}}$ from those of 
$\mbox{\boldmath $u$}_{\mathrm{best}}$ in the case of PG 1115+080. 
We use the time delays derived by \citet{barkana97}.
The errors of $a_0$ and $h$ of $\mbox{\boldmath $u$}_{\mathrm{best}}$
are 0.01 and 0.1, respectively. The isothermal model corresponds
to $\beta=1$.}
\label{fig:linear}
\end{figure}

The linear method that we used in this work is as follows. First, we choose a
model of Fourier expansion of $F(\theta)$ for a fixed value of $\beta$.
Second, for the observed image positions and the relative time delays, we
solve a set of linear equations [equations (\ref{eq:lenseq}) and (\ref{eq:tij})] 
\begin{eqnarray}
 A \mbox{\boldmath $u$}= \mbox{\boldmath $v$},
\end{eqnarray}
where $N \times M$ matrix $A$ and $N$ dimensional vector \mbox{\boldmath
$v$} are functions of observable quantities, such as the image positions
and the relative time delays, to obtain an initial guess of the
parameters $\mbox{\boldmath $u$}_{\mathrm{linear}}$ from which we start to
search a minimum of the $\chi^2$. Explicitly, the matrix $A$ is
expressed as 
\begin{eqnarray}
\left(
\begin{array}{cccccccc}
a_{11} & a_{21} & \alpha_{01} & \alpha_{m1} &
\beta_{m1} & \gamma_{11} & \gamma_{21} & 0\\
a'_{11} & a'_{21} & 0 & \alpha'_{m1} &
\beta'_{m1} & \gamma'_{11} & \gamma'_{21} & 0\\
. & . & . & . & . & . & . & 0\\
\Delta x_{ij} & \Delta y_{ij} & \hat{\alpha}_{0ij} &
\hat{\alpha}_{mij} & \hat{\beta}_{mij} &
\hat{\gamma}_{1ij} & \hat{\gamma}_{2ij} & T \\
. & . & . & . & . & . & . & .
\end{array}
\right),
\end{eqnarray}
where
$a_{1i}=\cos \theta_i$,
$a_{2i}=\sin \theta_i$,
$\alpha_{0i}=\beta r_i^{\beta-1}$,
$\alpha_{mi}=\beta r_i^{\beta-1}\cos m\theta_i$,
$\beta_{mi}=\beta r_i^{\beta-1}\sin m\theta_i$,
$\gamma_{1i}=r_i \cos 2\theta_i$,
$\gamma_{2i}=r_i \sin 2\theta_i$,
$a'_{1i}=-\sin \theta_i$,
$a'_{2i}=\cos \theta_i$,
$\alpha'_{mi}=-m r_i^{\beta-1}\sin m\theta_i$,
$\beta'_{mi}=m r_i^{\beta-1}\cos m\theta_i$,
$\gamma'_{1i}=-r_i\sin 2\theta_i$,
$\gamma'_{2i}=r_i \cos 2\theta_i$,
$\Delta x_{ij}=x_i-x_j$,
$\Delta y_{ij}=y_i-y_j$,
$\hat{\alpha}_{0ij}=r_i^{\beta}-r_j^{\beta}$,
$\hat{\alpha}_{mij}=r_i^{\beta}\cos m\theta_i
-r_j^{\beta}\cos m\theta_j$,
$\hat{\beta}_{mij}=r_i^{\beta}\sin m\theta_i
-r_j^{\beta}\sin m\theta_j$,
$\hat{\gamma}_{1ij}=\frac{1}{2}(r_i^2\cos 2\theta_i
-r_j^2\cos 2\theta_j$),
$\hat{\gamma}_{2ij}=\frac{1}{2}(r_i^2\sin 2\theta_i
-r_j^2\sin 2\theta_j$), and
$T=D^{-1} \Delta t_{ij}$ ($[1+z_d]D_{\mathrm{d}}D_{\mathrm{s}}/cD_{\mathrm{ds}} \equiv D h^{-1}$),
respectively. The elements of $(2i-1)$th and $(2i)$th lines are from the
lens equations of an image $i$. The elements of lines that are below
$(2n)$th line are ones of observed time delay for an $n$ imaged system.
The vector $\mbox{\boldmath $u$}$ is given as 
\begin{eqnarray}
 \mbox{\boldmath $u$}^T=(x_{\mathrm{s}}, y_{\mathrm{s}}, a_0, a_m, b_m,
\gamma_1, \gamma_2,h).
\end{eqnarray}
Finally, the vector $\mbox{\boldmath $v$}$ is expressed as
\begin{eqnarray}
 \mbox{\boldmath $v$}^T=(r_1, 0, ..., \frac{1}{2}(r_i^2-r_j^2), ...).
\end{eqnarray}
Usually, the number of linear equations $N$ (at most $11$ for a
quadruple lens system) should be equal or larger than that of linear
model parameters $M$. However, we can truncate some of the linear equations
so that the number of linear equations is the same as that of the
parameters, $N=M$. Thus, the parameters derived from the linear method is
$\mbox{\boldmath $u$}_{\mathrm{linear}}=A^{-1}\mbox{\boldmath $v$}$. We find
that this initial guess is close to the actual minimum of the $\chi^2$,
provided that the observed values have sufficiently small errors.
Especially, the Hubble constant, one of the most interesting parameters
cosmologically, of $\mbox{\boldmath $u$}_{\mathrm{linear}}$ coincides with
that of $\mbox{\boldmath $u$}_{\mathrm{best}}$ within $1\sigma$
uncertainties. Figure \ref{fig:linear} shows the result of the linear
method in PG 1115+080, the normalized deviations of $a_0$ and $h$ of
$\mbox{\boldmath $u$}_{\mathrm{linear}}$ from those of $\mbox{\boldmath
$u$}_{\mathrm{best}}$ within $1\sigma$ errors as a function of $\beta$
($0.01$ for $a_0$ and $0.1$ for $h$). 
Therefore, this method is useful enough to estimate the Hubble constant
accurately and quickly without solving nonlinear equations.

Our linear method is complementary to that of \citet{evansw03} in the
sense that our method is applicable to arbitrary values of  $\beta$ and
can derive the external shear as well (the essence of theirs
is the fact that for $\beta=1$ the magnification factor is expressed
as a linear equation with respect to the model parameter,
\begin{eqnarray}
 \mu (\mbox{\boldmath $x$})&=&\left\{ (1-\kappa)^2
      -\left[|\gamma|^2+\kappa^2
      -\frac{4\kappa}{r^2}\psi_{\mathrm{ex}}\right] \right\}^{-1} \\
&=& \left\{1-2\kappa(1-\gamma_1 \cos 2 \theta-\gamma_2 \sin 2 \theta)
      - |\gamma|^2 \right\}^{-1},
\end{eqnarray}
when the external shear is assumed).
The two methods can be combined in the case of $\beta=1$; first one
estimates the external shear with the former and then applies the
latter to the considered system in order to
reproduce the flux ratios {\it without} assuming the external shear.
However, this combined method sometimes suffers from results of
unphysical galaxy shapes, such as cross-like ones.

\end{document}